\documentclass[12pt,preprint]{iopart}

\usepackage{iopams}
\usepackage{indentfirst}
\usepackage{graphicx}
\usepackage{hyperref} 

\begin{document}

\title[Superallowed $0^+ \rightarrow 0^+$ beta decay studies at GANIL]{Superallowed $0^+ \rightarrow 0^+$ $\beta$ decay studies at GANIL and upcoming opportunities with DESIR and S$^3$-LEB}
\author{B.~M. Rebeiro$^1$, J.-C. Thomas$^1$, B. Blank$^2$ }
\address{$^1$ Grand Acc\'{e}l\'{e}rateur National d'Ions Lourds (GANIL), CEA/DRF-CNRS/IN2P3, Boulevard Henri Becquerel, F-14076 Caen, France}
\address{$^2$ Universit\'{e} de Bordeaux, CNRS/IN2P3, LP2I Bordeaux, UMR 5797, 33170 Gradignan, France}
\ead{bernadette.rebeiro@ganil.fr}

\vspace{8pt}

\begin{abstract}
Corrected transition rates ($\mathcal{F}t^{0^+ \rightarrow 0^+}$) of $0^+ \rightarrow 0^+$ superallowed $\beta$ decays currently give the most precise value of $V_{ud}$, the dominant term of the Cabibbo-Kobayashi-Maskawa (CKM) quark mixing matrix. By setting stringent constrains on the CKM unitarity, these decays allow  probing physics beyond the Standard Model in the electroweak sector. A recent global reevaluation of the $\mathcal{F}t^{0^+ \rightarrow 0^+}$ values has indicated a violation of CKM unitarity prompting reassessment of the theoretical radiative and isospin symmetry breaking corrections applied on the experimental transition rates $ft$. In this article we briefly discuss this current situation and the experimental program at GANIL geared towards constraining isospin symmetry breaking corrections. We conclude by presenting the opportunities that will be available at DESIR and S$^3$-LEB, the upcoming low-energy radioactive ion beam facilities at GANIL.
\end{abstract}
%
%
%
%
\section{Introduction}
\nopagebreak
Precision measurements of observables in superallowed (SA) Fermi $\beta$ transitions offer excellent probes to hunt for signatures of physics beyond the Standard Model (BSM)~\cite{GonzalezAlonso2019}. These decays proceed from the $J^\pi = 0^+$ ground-state of the parent nucleus to the $J^\pi = 0^+$ isobaric analog state in the daughter. In these decays, only the vector part of the weak interaction contributes. Consequently, the hypothesis of a conserved vector current (CVC)~\cite{Feynman1958} predicts that the transition rate ($ft$ value) remains constant across all SA decays for a given isospin, irrespective of the parent-daughter pair. This hypothesis has been tested and verified to a precision of $5\times 10^{-4}$~\cite{Hardy2020} once individual $ft$ values are corrected for nuclear structure and radiative effects (denoted $\mathcal{F}t$). The validation of CVC thus allows to derive $V_{ud}$, the dominant top-row element of the Cabibbo-Kobayashi-Maskawa (CKM) matrix, that describes the relation between quark flavor eigenstates and mass eigenstates. The unitarity of this CKM matrix is fundamental to the current version of the Standard Model and is tested to the highest precision using the first-row  elements. Being the largest and the most precise among the three elements, precision on $V_{ud}$ is key to performing sensitive tests of CKM unitarity. In fact, the highest precision on $V_{ud}$ currently comes from $\mathcal{F}t$ values of superallowed $0^+ \rightarrow 0^+$ beta decays ($\mathcal{F}t^{0^+ \rightarrow0^+}$)~\cite{Hardy2020,Falkowski2023}. 
Until recently, CKM unitarity was respected, however, a recent global re-evaluation of the $\mathcal{F}t^{0^+ \rightarrow0^+}$ values  has indicated a tension with unitarity at the $2\sigma$ level~\cite{Hardy2020,Falkowski2023}. Deviation from unitarity raises questions on the maximal violation of parity in weak interactions, thus opening doors to possible existence of exotic scalar $(S)$ and/or tensor $(T)$ currents in the electroweak sector~\cite{Hardy2020,Falkowski2023}. Another effect of relaxing maximal parity violation is the possibility of right-handed interactions via the $V+A$ coupling~\cite{Hardy2020}. Other scenarios that have been proposed to explain the non-unitarity of the CKM matrix, such as additional neutral gauge bosons, another generation of quarks, supersymmetry, etc, have been explored in Ref~\cite{Towner2010,Cirigliano2023}, to quote a few.

\begin{figure}[t]
\hspace*{-1.2 em}
\includegraphics[scale=0.55]{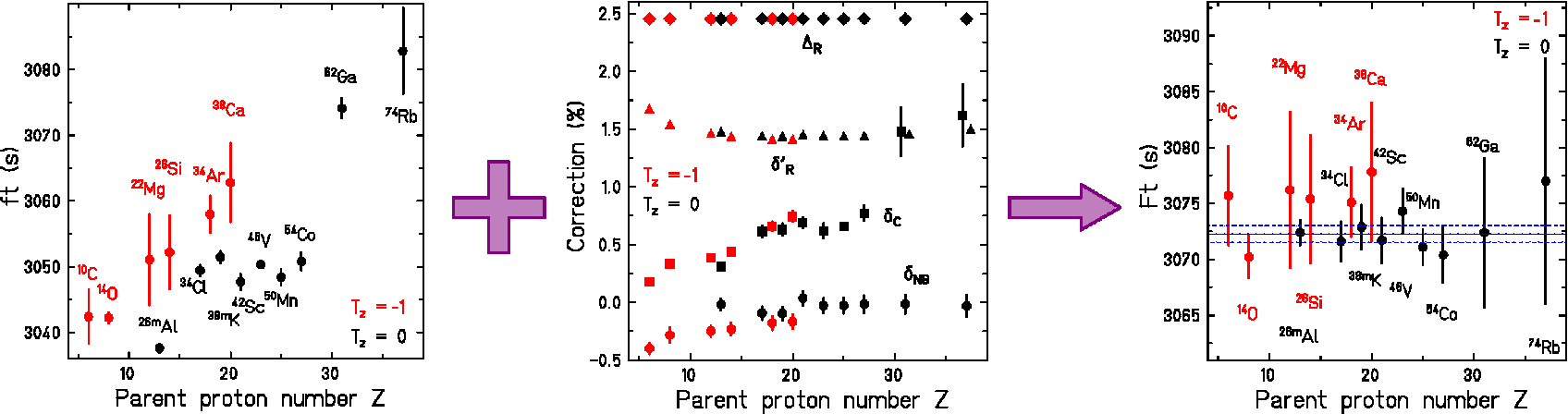}
 \caption{{\label{Fig:SA_ft_Corrections}}Transformation of nucleus-dependent measured $ft^{0^+ \rightarrow 0^+}$ values (left panel) to nuclear medium-independent $\mathcal{F}t^{0^+ \rightarrow 0^+}$ values (right panel) when radiative and isospin symmetry breaking corrections (central panel) are applied (see Eq.~\ref{Eq:Ft_Vud}). The two colors indicate the SA $\beta$ decay pairs, where the parent nucleus is either the even-even $T_Z = -1 $ (red points) or the odd-odd  $T_Z = 0 $ (black points) member of the $T=1$ isospin multiplet. Figure created with data from Ref~\cite{Hardy2020}.}
\end{figure}

Observables in SA $0^+ \rightarrow 0^+$ $\beta$ decays, namely branching ratio (BR), half-life ($t_{1/2}$), and Q-value, are used to determine the experimental $ft^{0^+ \rightarrow 0^+}$ value for the decay~\cite{Falkowski2023}. Individual $ft^{0^+ \rightarrow 0^+}$ values, shown in the left panel of Fig.~\ref{Fig:SA_ft_Corrections}, are corrected for nuclear medium dependent isospin symmetry breaking (ISB) and radiative effects~\cite{Hardy2020} to obtain the nucleus independent $\mathcal{F}t^{0^+ \rightarrow 0^+}$ values shown in the right panel of Fig.~\ref{Fig:SA_ft_Corrections}. The average of these $\mathcal{F}t^{0^+ \rightarrow 0^+}$ values is then used to derive $V_{ud}$, 
\begin{equation}
 \mathcal{F}t^{0^+ \rightarrow 0^+} = ft^{0^+ \rightarrow 0^+} (1+ \delta^{'}_{R} ) \left(1+ \delta_{NS}-\delta_C \right) = \frac{K}{2 V_{ud}^2 G_F^2 \left( 1+\Delta_R \right)}.
 \label{Eq:Ft_Vud}
\end{equation}
Here, the right hand side forms a constant with $K~=~8120.2787(11) \times 10^{-10}$ GeV$^{-4}$s and the nucleus-independent radiative correction $\Delta_R$, which is the same for all $0^+ \rightarrow 0^+$ SA $\beta$ transitions~\cite{Hardy2015}.
In the middle term of Eq.~\ref{Eq:Ft_Vud}, $\delta^{'}_{R}$ and $\delta_{NS}$ are radiative corrections, and $\delta_C$ is the isospin symmetry breaking correction. $\delta^{'}_{R}$ is independent of the details of the nuclear structure but depends on the atomic number Z of the daughter and the maximum energy of the beta particle emitted in the decay~\cite{Falkowski2023}. Computation of $\delta_{NS}$ on the contrary requires detailed description of the nuclear structure of both the initial and final nuclei, details of which can be found in Ref~\cite{Hardy2020}. As the average of $\mathcal{F}t^{0^+ \rightarrow 0^+}$ defines the absolute value of $V_{ud}$, it is absolutely necessary to  constrain these correction terms.

\begin{figure}[t]
  \centering
 \includegraphics[scale=0.15]{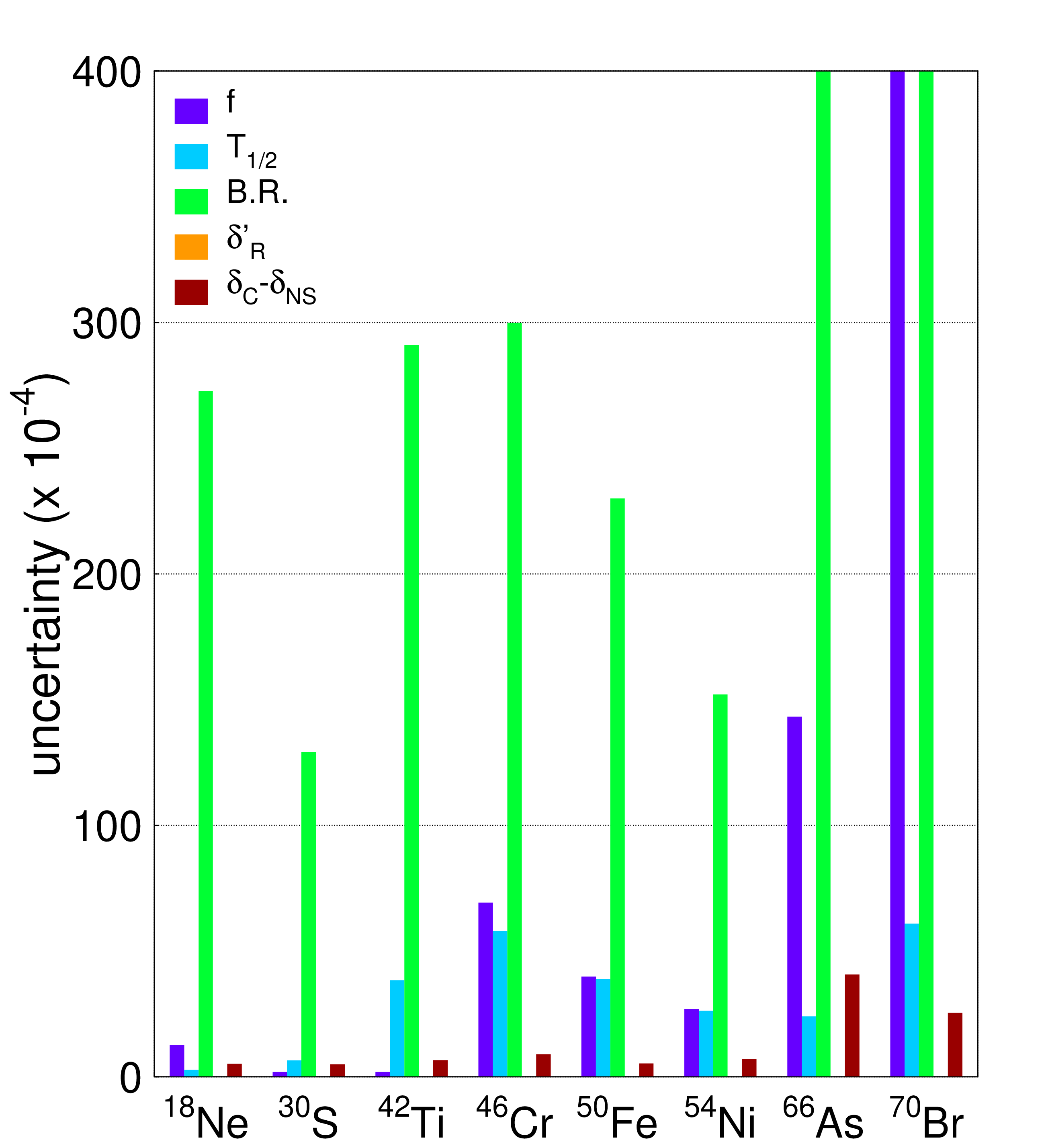}
 \includegraphics[scale=0.96]{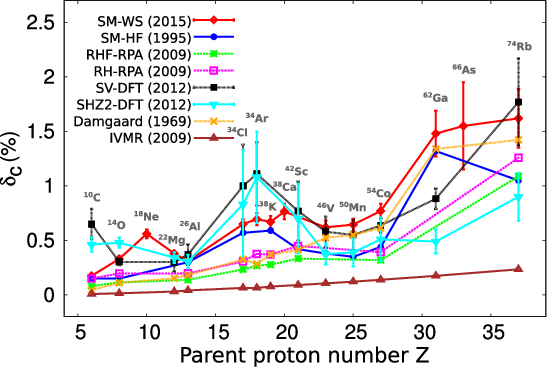}
 \caption{{\label{Fig:Uncertainty_ISBCorrections}}
{\bf Left panel}: Uncertainty contribution to the corrected $\mathcal{F}t^{0^+ \rightarrow 0^+ }$ for the eight less precise cases of superallowed $\beta$ emitters. Figure created with data from Ref.~\cite{Hardy2020}.  {\bf Right panel}: ISB corrections for  $0^+ \rightarrow 0^+$ SA decays from different models: SM-WS (2015) from Ref~\cite{Hardy2020}, SM-HF (1995) from Ref~\cite{Ormand1995}, RHF-RPA (2009) and RH-RPA (2009) from Ref~\cite{Liang2009}, SV-DFT (2012) and SHZ2-DFT (2012) from Ref~\cite{Satula2012}, Damgaard from Ref~\cite{Damgaard1969} and IVMR (2009) from Ref~\cite{Auerbach2009}. Figure adapted from Ref~\cite{Smirnova2023}.}
\end{figure}

The nuclear medium-dependent correction to $ft^{0^+ \rightarrow 0^+}$, as observable in Fig.~\ref{Fig:SA_ft_Corrections}, range from  $\sim -0.5\%$ to $ \sim 2\%$, and among the three, $\delta_C$ increases strongly with the atomic charge $Z$. This is natural, since the Coulomb interaction grows stronger with $Z$, however, this also enhances isospin mixing and symmetry breaking effects, consequently affecting the average $\mathcal{F}t^{0^+ \rightarrow 0^+}$ values.
Several different theoretical approaches have been used to calculate $\delta_C$~\cite{Hardy2020,Ormand1995,Liang2009,Satula2012,Satula2016,Damgaard1969,Auerbach2009}, and presented in the right panel of Fig.~\ref{Fig:Uncertainty_ISBCorrections}. As evident, there are large variations in $\delta_C$ when calculated using different models.
While this correction itself is a few percent, the change in $V_{ud}$ due to variation in $\delta_C$ is non-negligible. For instance, the authors of Ref~\cite{Grinyer2010} pointed out that when $\delta_C$ was recalculated in the shell-model formalism to include contributions from core orbitals~\cite{Towner2008}, the value of $V_{ud}$ increased by 1.8$\sigma$. They further discuss how the choice of model used to calculate $\delta_C$ affects the value of $V_{ud}$, mentioning a 2.5$\sigma$ deviation from CKM unitarity resulting from a sixfold increase in $V_{ud}$~\cite{Grinyer2010}. At present, only the shell model calculation using the Wood-Saxon wave function (annotated as SM-WS in Fig.~\ref{Fig:Uncertainty_ISBCorrections}) is used to correct the $ft^{0^+ \rightarrow 0^+}$ values for ISB effects~\cite{Hardy2020}. Constraining these ISB corrections requires high precision experimental data where these corrections are most sensitive, for e.g., near shell closures, where variations between models is large, or for high $Z$ pairs where the correction itself is large.

\begin{figure}[t]
\includegraphics[scale=0.4]{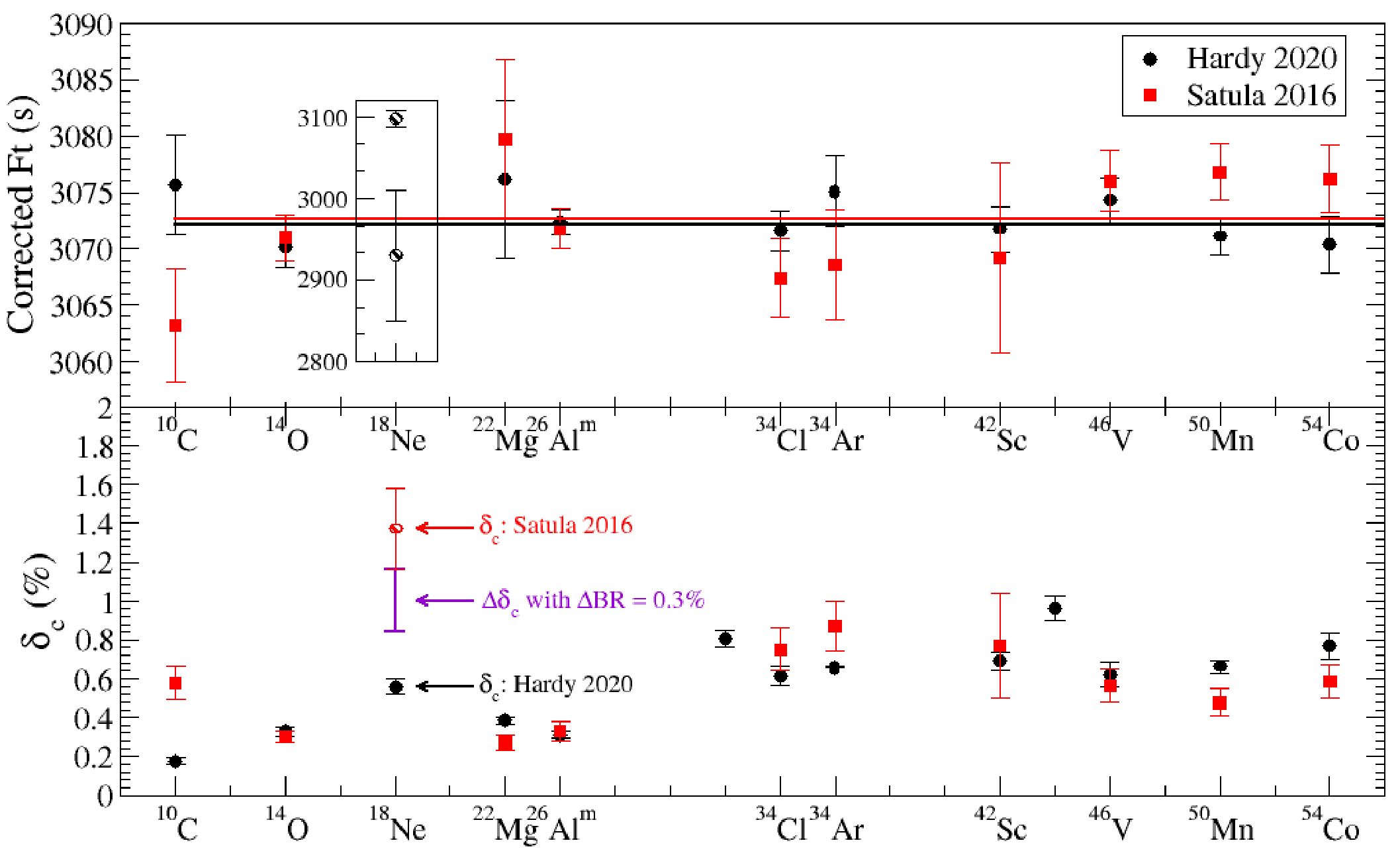}
\caption{{\label{Fig:Ne18_deltaC}
{\bf Top panel}: $\mathcal{F}t^{0^+ \rightarrow 0^+ }$ obtained using two different nuclear structure models that calculate the ISB corrections for $ft^{0^+ \rightarrow 0^+}$. Black data points  correspond to shell model based corrections from Ref~\cite{Hardy2020}, referred to as SM-WS in Fig.~\ref{Fig:Uncertainty_ISBCorrections}. Red data points correspond to density functional based corrections of Ref~\cite{Satula2016}, referred to as SV-DFT in Fig.~\ref{Fig:Uncertainty_ISBCorrections}. The two straight lines represent the average $\mathcal{F}t^{0^+ \rightarrow 0^+ }$ calculated using the 15 cases shown in Fig.~\ref{Fig:Uncertainty_ISBCorrections}. The inset displays the $\mathcal{F}t^{0^+ \rightarrow 0^+ }$ value for $^{18}$Ne determined using existing experimental data and the ISB corrections of Ref~\cite{Hardy2020}. \\
{\bf Bottom panel}: ISB corrections $\delta_C$ taken from Ref~\cite{Hardy2020} and \cite{Satula2016}. Assuming $\delta_C=1$, we apply the CVC filter described in Ref~\cite{Hardy2020} and propagate the uncertainties from the $^{18}$Ne $ft^{0^+ \rightarrow 0^+}$ value, the average of the $\mathcal{F}t^{0^+ \rightarrow 0^+}$ values for the 15 most precise SA $\beta$ emitters, as well as the theoretical uncertainty in $\delta^{'}_{R}$ and $\delta_{NS}$. This yields the purple error bar, that demonstrates the potential to distinguish between the two models if the branching ratio for $^{18}$Ne can be measured with a precision of 0.3\%.}}
\end{figure}

Currently, the $ft^{0^+ \rightarrow 0^+}$ values are measured for 23 SA $\beta$ emitters, ranging from $^{10}$C $(Z=6)$ to $^{74}$Rb $(Z=37)$~\cite{Hardy2020}. However, only 15 of these, shown in Fig.~\ref{Fig:SA_ft_Corrections}, are used in the evaluation of $V_{ud}$ because their $ft^{0^+ \rightarrow 0^+}$ values have the required precision of $0.3\%$ or better~\cite{Hardy2020}. For the remaining eight cases, that are shown in the left panel of Fig.~\ref{Fig:Uncertainty_ISBCorrections}, the uncertainty is dominated by those on the BR and Q-value (used to determine $f$ in Fig~\ref{Fig:Uncertainty_ISBCorrections}).
To effectively constrain the ISB corrections, a precision of $0.3\%$ or better is required on the experimental $ft$ values~\cite{Hardy2020}. 
In Fig.~\ref{Fig:Ne18_deltaC}, we demonstrate using $^{18}$Ne, how a precision of $0.3\%$ on the SA branching ratio for this decay, can help distinguish between two advanced models.

\section{Experimental program}
At GANIL, an experimental program is currently underway aiming towards constraining ISB corrections by performing high-precision measurements of the half-life and branching ratio of the eight less-precisely known SA $\beta$ emitters. Of these eight nuclei, the first six (shown in Fig.~\ref{Fig:Uncertainty_ISBCorrections}) are $T_Z = -1$ parents, where the $\delta_C$ corrections are larger than those in $\beta$ decay pairs with $T_Z = 0$ parents of the same mass. Furthermore, for the SA parent nuclei $^{18}$Ne, $^{30}$S and $^{42}$Ti - which, like $^{34}$Ar, exhibit closed neutron shell configurations - some models predict a systematically larger value of $\delta_C$ at shell-closures. Others, however, show no such variations (e.g., brown triangles in Fig.~\ref{Fig:Uncertainty_ISBCorrections}), leading to larger discrepancies between different models. Additionally, as the atomic number increases, the magnitude of $\delta_C$ increases, and as one can observe for the three heaviest SA emitters in Fig.~\ref{Fig:Uncertainty_ISBCorrections}, these corrections can range from approximately 0.1\% to 1.8\%. Given the sensitivity of these candidates for evaluating $\delta_C$, the experimental program at GANIL focuses on improving the precision of these eight known SA emitters and extending this study to the $N=Z$ SA emitters beyond $^{74}$Rb to test the $\delta_C$ models for larger variations in $Z$.

{\subsection{\label{PastExpts} Past superallowed beta decay studies at GANIL }}
As part of this effort, measurements were conducted at GANIL to determine the half-life and SA branching ratio of $^{38}$Ca~\cite{BlankEPJ2015} and $^{30}$S~\cite{S30_Aouadi2017} using the in-flight fragment separator LISE~\cite{LISE3}. In these experiments, the detection setup consisted of plastic scintillators for detecting $\beta$ radiation, monitoring beam purity, and implantation onto a moving tape-drive system, as well as HPGe detectors for measuring the de-exciting $\gamma$ radiations. The tape-drive system allowed separating the implantion and decay positions, which is essential to measure the decay in a background-free environment and to remove activity from long-lived beam contaminants and unstable decay products. To determine the branching ratio, a High Purity Germanium (HPGe) detector was used in coincidence with a thin plastic scintillator. The HPGe detector was characterized in efficiency at the 0.2\% level when placed 15~cm from the decay point~\cite{Blank2015,Blank2020}. Such a precise characterization is critical for achieving the 0.3\% precision on the BR required for these measurements. The incident beam intensity and cycle times for these measurements were also optimized to collect enough statistics to meet the precision requirements. The beam was implanted for $3t_{1/2}$ and decay time ranged from $ 3t_{1/2}$ for branching ratio measurements to $10t_{1/2} $ - $ 20t_{1/2}$ for half-life measurements. The cycle time also included background measurements before implantation and after moving the activity into storage cassettes.

{\subsection{\label{UpcomingExpts} Upcoming superallowed beta decay studies at SPIRAL1 and LISE at GANIL}}
A similar experiment will be conducted at GANIL in June 2025 to measure the half-life and SA branching ratio in $^{18}$Ne~\cite{Ne18_Reb2023}. This measurement will be done using the radioactive $^{18}$Ne beam produced at SPIRAL1 (Syst\`{e}me de Production d'Ions Radioactifs Acc\'{e}l\'{e}r\'{e}s en Ligne phase 1)~\cite{Delahaye2020}, the Isotope Separation On-Line (ISOL) facility of GANIL. The currently published branching ratio~\cite{Hardy2020} for this decay has a precision of $\pm$2.7\%, which is used to determine the $^{18}$Ne $ft^{0^+ \rightarrow 0^+ }$ value. A second measurement~\cite{Haifa_Thesis}, published in a Ph.D. thesis, quotes a statistical uncertainty of $\pm 0.7\%$. This second measurement was done at GANIL but the final precision on the BR was limited by statistics and affected by rate-dependent effects.
The later arose from the use of peak-sensing, VXI-based analog-to-digital converters (ADCs) for measuring the branching ratio. At high counting rates, such ADCs preferentially register high-energy $\gamma$-rays, thus biasing the measured relative intensity. Furthermore, pulse pileup within the coding gate reduces the intensity derived for the strongest gamma rays. The lack of statistics also resulted due to tape breaks with the then newly commissioned moving tape system at the low-energy identification station at SPIRAL1~\cite{Grinyer2014}.

The upcoming $^{18}$Ne measurement~\cite{Ne18_Reb2023} will use a similar $\beta-\gamma$ detection setup as mentioned in section~\ref{PastExpts} above, a moving tape system, a plastic scintillator for $\beta$ detection and the same HPGe detector~\cite{Blank2015,Blank2020} mentioned above for $\gamma$ detection. To address rate-dependent effects encountered in the previous $^{18}$Ne measurement~\cite{Haifa_Thesis}, signals from the various detectors will be processed using the FASTER~\cite{FASTER} digital data acquisition (DAQ) system as opposed to the previously used VXI based analogue DAQ. FASTER is a triggerless digital DAQ that timestamps detected events with a precision of 2~ns. Furthermore, signal treament in FASTER requires less than $10\mu s$ compared to the VXI DAQ that requires several 10s of $\mu s$. This helps reduce the systematic deadtime making FASTER suitable for high-rate measurements. In addition, pulse shape analysis of the detector signals will allow to identify and accurately quantify pileup events.

\begin{figure}
\centering
\includegraphics[scale=0.45]{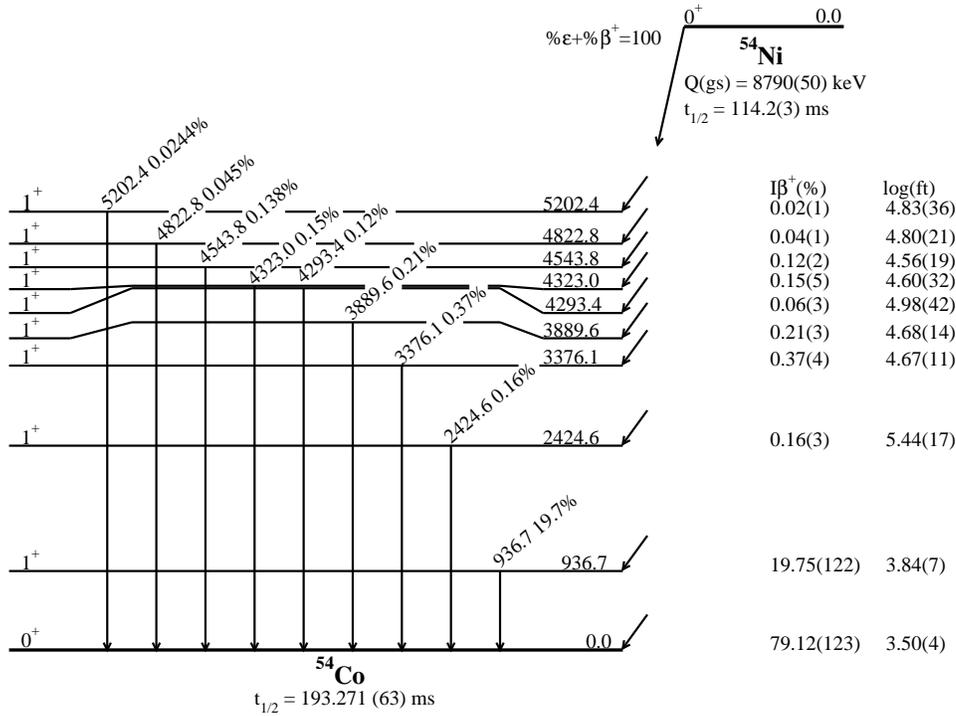}
\caption{{\label{Fig:54Ni_decay}
Known decay radiation scheme for the superallowed emitter $^{54}$Ni. Figure created with data retrieved from Refs~\cite{ENSDF_Ni54} and \cite{Molina2015}. }}
\end{figure}

In the near future, we will also conduct experiments at LISE to improve the precision on the SA branching ratio for $^{42}$Ti, $^{46}$Cr, $^{50}$Fe and $^{54}$Ni. Beyond $Z=20$, the neutron-deficient radioactive ion beams (RIBs) that can be produced via in-flight fragmentation at LISE are not pure enough (typical purity requirement is $>99\%$) to perform the high-precision measurements required for SA $\beta$ decay studies. The purity and consequently the RIB intensity is compromised due to the overlap of momentum distribution tails from the less exotic isotones, which are produced at much higher intensities in the fragmentation process. Additionally, these $T_Z=-1$ parents decay to daughter nuclei that have a significant number of energetically available $J^\pi = 1^+$ states. These $0^+ \rightarrow 1^+$ Gamow-Teller (GT) transitions split the $\beta$ decay strength as they compete with the $0^+ \rightarrow 0^+$ superallowed $\beta$ decay branch. In such cases, the branching ratio for the SA decay must be determined indirectly by measuring all the non-analog branches and subtracting their sum from the total $\beta$ decay,
\begin{equation}
 BR_{SA} = 1 - \Sigma_i BR_i .
 \label{Eq:SA_BR_Sum}
\end{equation}
In Eq.~\ref{Eq:SA_BR_Sum} above, $i$ runs over all non-SA $\beta$ decay branches, which need to be measured with sufficiently high precision to obtain the targeted $0.3\%$ precision on the SA decay branch. Some of these non-analog branches, however, are weaker and the de-exciting gamma radiations have high energies, as can be seen for the case for $^{54}$Ni in Fig.~\ref{Fig:54Ni_decay}. This sets up the stage for the ``Pandemonium  effect'' \cite{Hardy1977} when high resolution HPGe detectors are used to determine the branching ratios. This systematic effect arises from an underestimation of the $\beta$ decay strength at high energy levels, which occurs when weak $\gamma$ cascades or high-energy $\gamma$ radiations are not detected or are missed due to the low efﬁciency of such high resolution germanium detectors.

To address these issues at LISE, the SA parent will be implanted in a segmented charged-particle detector instead of a moving tape system. To determine the half-life
while avoiding $\beta$ radiations from contaminants, a suitable time correlation will be applied between the events corresponding to the implant position of the SA parent and the $\beta$ decay events. To address the Pandemonium effect, gamma detectors that perform as Total Absorption Spectrometers (TAS) such as the ones being developed within the (NA)$^2$STARS project~\cite{NA2STARSProp} will be used to determine the branching ratio. These TAS $\gamma$ detectors are large scintillators (e.g., NaI or BaF$_2$ ) that have high $\gamma$ detection efficiency and are used as calorimeters, covering a nearly $4\pi$ geometry to detect the entire cascade of $\gamma$ decays rather than individual $\gamma$ rays. To address contamination from radioactive daughters, the measurement cycles will be optimized for appropriate implantion and decay times with accurate forward and backwards time correlation gates between the implantation and $\beta - \gamma$ events. One such experiment is scheduled to be performed at LISE in 2026~\cite{Orrigo2023} and will use a double-sided silicon strip detector (DSSSD) for implantation, plastic scintillators for $\beta$ detection, and either the DTAS detector (18 NaI(Tl) crystals)~\cite{Guadilla2018} or the Rocinante detector (12 BaF$_2$ crystals)~\cite{Valencia2017} as a total absorption spectrometer. For SA $\beta$ decay studies, this setup will be complemented with the previouslly mentioned HPGe detector~\cite{Blank2015,Blank2020}.

{\subsection{\label{S3_DESIR} Superallowed beta decay studies beyond $^{54}$Ni at GANIL's upcoming low energy facilities DESIR and S$^3$-LEB}}

As the atomic number increases and we move further up the nuclear chart towards $^{98}$In, the SA emitters become more exotic, and based on the trend in Fig.~\ref{Fig:Uncertainty_ISBCorrections}, the isospin symmetry breaking corrections will increase further~\cite{Smirnova2023}. To test the robustness of the nuclear structure models that calculate ISB corrections, we need high-precision experimental data across the nuclear chart, especially covering large variations in the atomic number $Z$. Currently, the uncertainty in the branching ratio for the known heavier SA beta emitters, such as $^{66}$As or $^{70}$Br, is at least 10 times larger than the minimum prescribed limit~\cite{Hardy2020}. As mentioned earlier, the purity of the neutron-deficient nuclei beyond $^{54}$Ni, that can be produced at LISE via in-flight fragmentation is severely limited. In addition, the refractory nature of the heavier SA emitters limits their production at SPIRAL1 via the current ISOL technique. Thus, the studies of SA emitters from $^{66}$As to $^{98}$In, focusing on the $N=Z$, $T_Z=0$ cases, requires developments in RIB production and purification methods.

Several accelerator facilities worldwide use the ISOL technique to produce radioactive ions either by proton-induced fission and spallation (ISOLDE at CERN, SPES at INFN Legnaro, TRIUMF in Canada, ANL in  USA) and light-particle induced transfer reactions (JYFL in Finland). Others produce RIBs via in-flight projectile fragmentation (GSI in Germany, NSCL in USA) similar to LISE at GANIL or in-flight fission (RIKEN in Japan) which is better suited for producing neutron-rich RIBs. These techniques are not well suited to produce the heavier neutron-deficient superallowed emitters at the required intensity and purity. As an example, $^{62}$Ga beams are available at ISOLDE and TRIUMF with intensities on the order of $10^3$ pps~\cite{ISOLDEIntensity,TRIUMFIntensity}. For a reasonable measurement time that could be alloted at such high-demand experimental facilities, attaining a statistical precision below 1\% on the SA branching will be challenging. An alternative method to produce these neutron-deficient RIBs is by fusion-evaporation reactions which has been implemented at IGISOL in Finland and will be implemented at GANIL where yields larger than $10^4$~pps are expected for $^{62}$Ga and $^{74}$Rb~\cite{GANILChartbeams,GANILbeams}.

Significant efforts are currently underway at GANIL to produce neutron-deficient RIBs at SPIRAL1, where, as mentioned earlier, the ISOL technique is being used. In this context, the on-going TULIP project~\cite{Jardin2023,Bosquet2023} aims to produce RIBs at SPIRAL1 via fusion-evaporation reactions, which also favors the production of neutron-deficient nuclei. As part of this project, the production of $^{74-78}$Rb$^+$ isotopes was recently demonstrated~\cite{Chauveau2024} and production yields for neutron-deficient galium isotopes (including the SA emitter $^{62}$Ga) will be investigated in an upcoming beam test run~\cite{Jardin}. Furthermore, with the beam purification techniques that will be implemented at GANIL's upcoming low-energy facilities DESIR ({\bf{D}}\'{e}sint\'{e}gration, {\bf{E}}xcitation et {\bf{S}}tockage d'{\bf{I}}ons {\bf{R}}adioactifs)~\cite{DESIR} and S$^3$-LEB ({\bf{S}}uper {\bf{S}}eparator {\bf{S}}pectrometer's {\bf{L}}ow {\bf{E}}nergy {\bf{B}}ranch)~\cite{S3LEB}, high-precision measurements of SA decay observables for nuclei beyond $^{54}$Ni will become possible. In this direction, S$^3$ conducted its first commissioning run in November 2024, while DESIR is gearing up for its first commissioning run in 2027, setting up opportunities for unprecedented high-precision measurements at GANIL.

The Super Separator Spectrometer of GANIL will take intense stable beams from SPIRAL2~\cite{Orduz2022,Gales2010} and produce RIBs in-flight via the fusion evaporation method, which is favorable for producing neutron-deficient nuclei, like the SA emitters in this project. The cocktail of RIBs that is produced will be slowed down and neutralized in a gas cell~\cite{Romans2022,Ajayakumar2023}. The neutralized atoms will be laser-ionized to select and transport the RIB of interest towards user setups. S$^3$-LEB will also be coupled to a multi-reflection time-of-flight spectrometer called PILGRIM, which will have a mass resolution $m/\Delta m$ ranging from $1.5\times10^5$ to $5.2\times10^5$, depending on the beam emittance~\cite{Chauveau2016}. For $10^4$ radioactive ions, this would allow to measure the masses with a precision of few keVs, which is at the level required for SA beta decay measurements.

Studying SA $\beta$ decays at S$^3$-LEB, currently presents a few challenges, the resolution of which will require some developments. First, the current version of the S$^3$-LEB gas cell has an extraction time of around 500~ms~\cite{Manea2021} which is much longer compared to the half lives of the heavier SA emitters, e.g. $t_{1/2}$ for $^{70}$Br is $78.88\pm0.31$~ms~\cite{NNDC}.  Efforts are currently being made within the S$^3$-LEB collaboration to reduce the extraction time to $\sim50$~ms using a fast gas cell called FRIENDS3 (Fast Radioactive Ion Extraction and Neutralization Device for S3)~\cite{FRIENDS3_ANR}, that is currently under development.  The second challenge is the availability of appropriate laser ionization schemes, and laser spectroscopy studies are not done yet on $^{66}$As, $^{70}$Br, $^{83}$Tc and $^{90}$Rh~\cite{Yang2023}. The collaboration has a laser laboratory at GANIL called GISELE~\cite{Romans2023} which has the same laser systems as at S$^3$-LEB and is coupled to an atomic beam unit that can be used to produce the stable element of interest for improving existing or developing new laser ionization schemes.

\begin{figure}[h]
\centering{
 \includegraphics[width=0.95\textwidth]{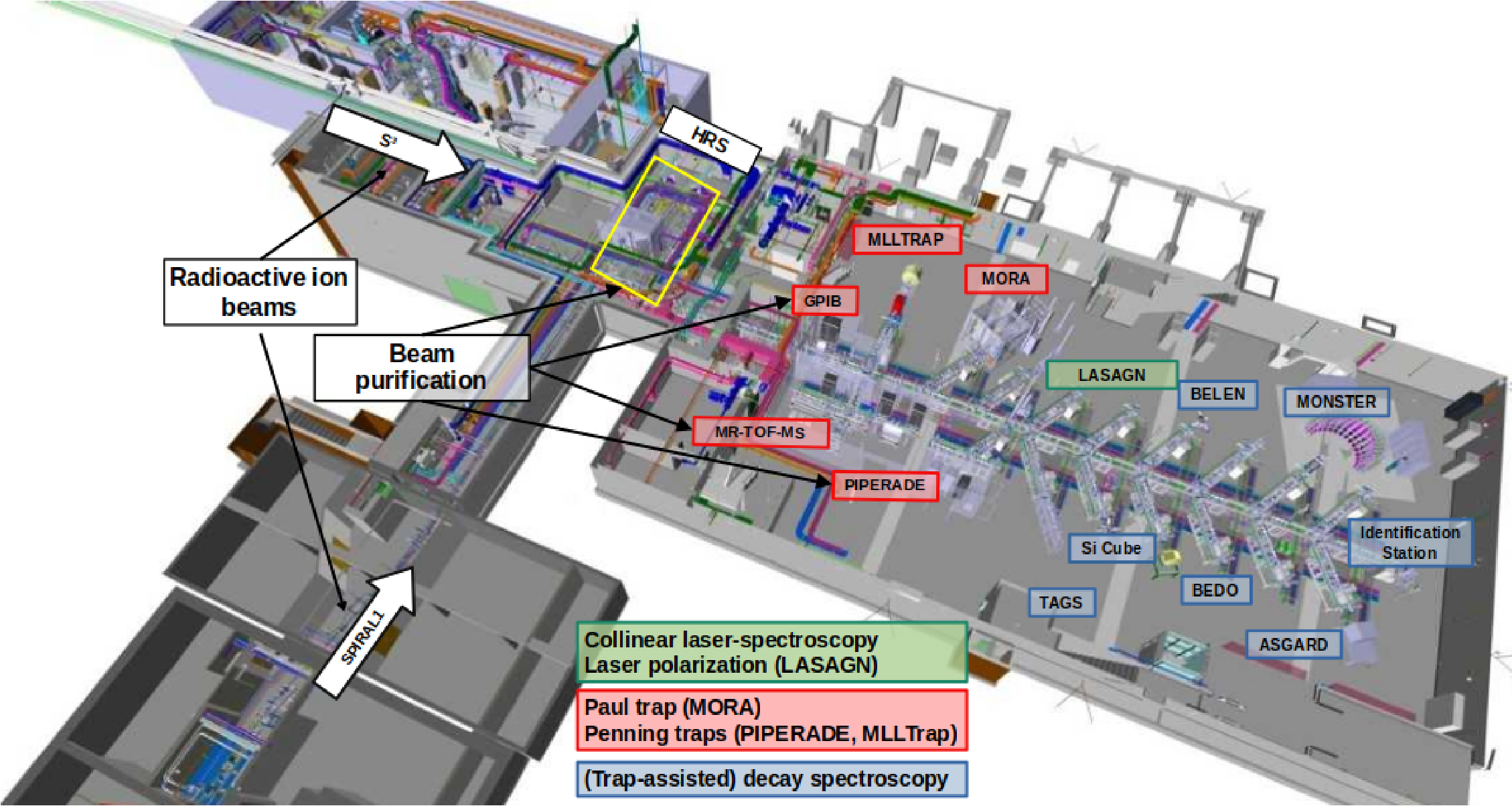}
\caption{{\label{Fig:DESIR_layout}
A schematic layout showing the tentatively proposed plan for the placement of different instruments in the future DESIR beam hall. Beams coming from SPIRAL1 and S$^3$ are shown with white solid arrows. All the decay spectroscopy (BESTIOL) associated setups are indicated in blue, ion trap devices (DETRAP) in red and laser spectroscopy devices (LUMIERE) are in green. Figure adapted from Ref~\cite{DESIR}}.}}
\end{figure}

The other facility for performing SA measurements at GANIL is DESIR, which will receive beams from both SPIRAL1 and SPIRAL2 via S$^3$-LEB. At DESIR, these beams will be purified first by a High Resolution Separator (HRS)~\cite{Michaud2023} which is designed to have a mass resolution of $ m /\Delta m  = 2\times 10^4 $. Further improvement in the beam quality will be done by cooling and bunching the beam in a gas filled Paul trap called General Purpose Ion Buncher (GPIB)~\cite{Gerbaux2023} which can be coupled to a double Penning trap called PIPERADE (PI\`{e}ges de PEnning pour les RAdionucl\'{e}ides \`{a} DEsir)~\cite{Ascher2021}. PIPERADE is currently being commissioned and offline tests have demonstrated a mass resolution of $ m /\Delta m  = 10^5 $~\cite{Ascher2021}. The ultimate aim is to attain a mass resolution of $ m /\Delta m  = 10^7 $. Since the beam purification at DESIR does not involve a gas cell or laser ionization, non-refractory nuclei beyond $^{54}$Ni (see Fig.~\ref{Fig:Uncertainty_ISBCorrections}) can be studied at DESIR with beams from SPIRAL1 obtained using the TULIP target ion source~\cite{Jardin2023,Bosquet2023}. 
The highly purified beam thus obtained will then be sent to different experimental setups placed in the DESIR hall, a layout of which is shown in Fig.~\ref{Fig:DESIR_layout}.

As part of DESIR's BESTIOL program, a dedicated decay spectroscopy station equipped with a moving tape drive system, charged particle detectors, and gamma detectors will be provided for decay studies. Additionally there will be provisions for users to set up their own detection systems that can be coupled to a mobile tape transport system. The availability of high purity beams, and experimental setups similar to those described in sections~\ref{PastExpts} and~\ref{UpcomingExpts},  make DESIR an excellent facility for studying the half-life and branching ratio of superallowed beta decays.

Furthermore, within the DETRAP program, DESIR will host high-resolution mass separation and measurement devices, such as the HRS~\cite{Michaud2023}, PIPERADE~\cite{Ascher2021}, and, in the near future, a Multi-Reflection Time-of-Flight Mass Spectrometer (MR-TOF-MS) based on the PILGRIM design~\cite{Chauveau2016}. These devices will have the capability to perform mass measurements with the precision required for SA $\beta$ decay measurements.
In addition, through its LUMIERE program, DESIR will also offer opportunities to perform laser spectroscopy of RIBs or conduct $\beta$-decay studies on laser-polarized RIBs. With multiple setups DESIR will serve as a low-energy, high-precision facility, where all the SA observables can be measured with the same primary radioactive beam.

\vspace*{1 em}

\section{Concluding remarks}
Search for physics beyond the Standard Model is currently the driver for several large-scale research programs worldwide. Precision measurements in nuclear $\beta$ decays have proved to be crucial probes for such investigations, particularly in determining $V_{ud}$ and testing the unitarity of the CKM matrix.
With the recent tensions in CKM unitarity, nuclear structure-dependent ISB corrections applied to $ft^{0^+ \rightarrow 0^+}$ have come under increased scrutiny, especially since a single model is used to determine $\mathcal{F}t^{0^+ \rightarrow 0^+}$ and consequently $V_{ud}$.  This has highlighted the need for improving the precision on $ft^{0^+ \rightarrow 0^+}$ values for some of the $T_Z = -1$ parents which are particularly sensitive to the ISB corrections. Additionally the sensitivity of $\delta_C$ to $Z$ urges studying SA $\beta$ decays beyond $^{74}$Rb to test the models against larger variations in $Z$.

In this article, we presented the on-going experimental program at GANIL, which aims to constrain ISB corrections through high-precision measurements of observables in SA $\beta$ decays. We also discussed the current limitations at GANIL for studying this decay mode in nuclei beyond $^{54}$Ni, and the opportunities that the upcoming DESIR and S$^3$-LEB facilities, along with the TULIP RIB development project, will provide for extending these studies towards $^{98}$In along the $N=Z$ path.

While in the current article we limited our discussions to SA $0^+ \rightarrow 0^+$ Fermi $\beta$ decays, a similar effort is ongoing within the French research community to determine $V_{ud}$ from SA mirror decays~\cite{Falkowski2021}.
%
%
Beyond CKM unitarity tests, the search for BSM physics also includes looking for the existence of scalar or tensor currents, which go beyond the vector-axial vector (V-A) nature of weak interactions. The search for such exotic currents is carried out by precisely measuring the $\beta$ energy spectrum or by performing correlation measurements. At GANIL, within the framework of the b-STILED ({\bf{b}}: improved {\bf{S}}earch for {\bf{T}}ensor {\bf{I}}nteractions in nuc{\bf{L}}ear b{\bf{E}}ta {\bf{D}}ecay) project, experiments were conducted at LISE and SPIRAL1 to study the $\beta$ energy spectrum in the decay of $^6$He~\cite{Kanafani2022}.
{\bf{M}}atter’s {\bf{O}}rigin from {\bf{R}}adio{\bf{A}}ctivity (MORA) is another high-precision experiment that will measure the {\it{D correlation}} in the decay of trapped and laser-polarized $^{23}$Mg$^+ $ and $^{39}$Ca$^+ $ ions. The MORA experiment~\cite{Delahaye2019,Goyal2025} is searching for CP-violating interactions in the V-A framework of the Standard Model and will perform its Phase II measurements at DESIR within the DETRAP program.
The ASGARD ({\bf{A}}luminium {\bf{S}}uperconducting {\bf{G}}rid {\bf{A}}ssembly for {\bf{R}}adiation {\bf{D}}etection) project~\cite{ASGARDKanafani} will measure the recoil energy of the daughter nucleus to determine the $\beta-\nu$ angular correlation, which is required to determine $V_{ud}$ from SA mirror $\beta$ decays. ASGARD will conduct these experiments at DESIR (and other ISOL facilities) and will also search for scalar and tensor currents by comparing electron capture and $\beta^+$ decay rates.
WISArD ({\bf{W}}eak {\bf{I}}nteraction {\bf{S}}tudies with $^{32}${\bf{Ar}} {\bf{D}}ecay) is another high-precision experiment investigating exotic scalar and tensor currents by studying the $\beta-\nu$ angular correlation of $^{32}$Ar~\cite{WISARD2019,WISARD2024}. The experiment focuses on detecting beta-delayed protons from the daughter nucleus $^{32}$Cl. Additionally, they plan to constrain the Fierz interference term by measuring beta-decay spectrum shape. While WISArD is currently setup at ISOLDE, the collaboration plans to replicate the setup to perform similar measurements on $^{20}$Mg at DESIR~\cite{WISARD2024}.

With several ambitious research programs, advanced experimental techniques and upcoming facilities, GANIL is in a unique and competitive position globally to explore the frontiers of physics and set stringent limits on the search for BSM physics.

\section{Acknowledgements}
Several projects mentioned in the article were supported by Agence Nationale de la Recherche, France through ANR-11-EQPX-0012 and ANR-10-EQPX-46 and the FEDER (Fonds Europeen de Developpement Economique et Regional) and CPER (Contrat Plan Etat R\'{e}gion) programs. A complete list of contributing funding agencies can be found in Refs.~\cite{Ajayakumar2023,Michaud2023,Gerbaux2023,Ascher2021}.

\section{Conflict of Interest}
The authors declare no conflict of interest.

\section*{References}

\end{document}